\documentclass[aps,prd,10pt,onecolumn]{revtex4}
\usepackage{amsmath}
\usepackage{amssymb,epsf}
\usepackage{latexsym}

\begin{document}

\title{Quasinormal modes of black holes in Weyl gravity:\\
Electromagnetic and gravitational perturbations}
\author{Mehrab Momennia$^{1}$\footnote{email address: m.momennia@shirazu.ac.ir} and Seyed Hossein
Hendi$^{1,2}$\footnote{email address: hendi@shirazu.ac.ir}}
\affiliation{$^1$Physics Department and Biruni Observatory,
College of Sciences, Shiraz
University, Shiraz 71454, Iran\\
$^2$Canadian Quantum Research Center 204-3002 32 Ave Vernon, BC
V1T 2L7 Canada}

\begin{abstract}
The recent reported gravitational wave detection motivates one to
investigate the properties of different black hole models, especially their
behavior under (axial) gravitational perturbation. Here, we study the
quasinormal modes of black holes in Weyl gravity. We derive the master
equation describing the quasinormal radiation by using a relation between
the Schwarzschild-anti de Sitter black holes and Weyl solutions, and also
the conformal invariance property of the Weyl action. It will be observed
that the quasinormal mode spectra of the Weyl solutions deviate from those
of the Schwarzschild black hole due to the presence of an additional linear $%
r$-term in the metric function. We also consider the evolution of the
Maxwell field on the background spacetime and obtain the master equation of
electromagnetic perturbations. Then, we use the WKB approximation and
asymptotic iteration method to calculate the quasinormal frequencies.
Finally, the time evolution of modes is studied through the time-domain
integration of the master equation.
\end{abstract}

\maketitle

\section{Introduction}

From theoretical point of view, black hole is one of the most interesting
and important solutions to gravity theories. The existence of such highly
dense object in the cosmos has been also proved through the detection of
gravitational waves (GWs) of black hole binary mergers \cite{AbbottBH} by
the LIGO and VIRGO observatories and the captured image of the `shadow' of a
supermassive black hole by the Event Horizon Telescope collaboration \cite%
{Akiyama1,Akiyama4}. In this regard, it is interesting to investigate the
other kinds of black hole solutions such as ones which are constructed based
on higher curvature theories of gravity.

Weyl gravity, which its Lagrangian is defined by the square of the Weyl
tensor, is one of the successful and interesting theories in higher
derivative gravity scenario \cite{Bach,Buchdahl,Riegert,Mannheim,Pope}. This
model of gravity is invariant under local scale transformation of the metric
$g_{\mu \nu }(x)\rightarrow \Omega ^{2}(x)g_{\mu \nu }(x)$, and thus, it is
unique up to the choice of the matter source in order to keep the conformal
invariance property. The Weyl gravity suffers the Weyl ghost that it might
be possible to remove it under certain conditions \cite%
{MannheimDavidson,Bender,BenderMannheim,BenderMannheim2008,BenderMannheimPRL,Mannheim2013,MannheimPLB,Mannheim2018}%
. In addition, one can consider this theory of gravity as a suitable model
for constructing quantum gravity \cite{Bergshoeff,Wit}, since it is a
higher-curvature theory of gravity which is power-counting renormalizable
\cite{Stelle,Faria}.

From the viewpoint of high energy physics, it is shown that the Weyl gravity
arises from twister-string theory with both closed strings and gauge-singlet
open strings \cite{Witten}, and it appears as a counterterm in adS/CFT
calculations \cite{Tseytlin,Henningson,Balasubramanian}. In addition, this
theory can be examined as a possible UV completion of Einstein gravity \cite%
{Adler,Hooft}. It is worthwhile to mention that there is an equivalence
between Einstein gravity and Weyl gravity by considering the Neumann
boundary conditions \cite{Maldacena,Anastasiou}.

The discovery of GWs produced by the merger of compact binary objects \cite%
{AbbottBH,AbbottNS} added a totally different and new type of observations
to the traditional astronomy based on electromagnetic waves. The emitted GWs
contain a lot of information for fundamental physics and one can check the
validity of the alternative theories of general relativity, such as massive
gravity, scalar-tensor theory, and Weyl gravity. The signal of compact
binary merger can be decomposed into three phases \cite{Blanchet}. The first
stage of a binary system is the inspiral phase which the frequency and
amplitude of the signal chirp with time. During this phase, the signal is
universal that just depends on the masses and the spins of compact objects
and does not depend on the nature of source. The post-Newtonian
approximation is a powerful tool for describing the inspiral phase \cite%
{Blanchet}. The second stage is the merger phase and occurs after the
inspiral phase with a\ rapid collapse of the two objects to form a black
hole. The amplitude of GWs has a peak at this time and numerical relativity
is used to compute the merger phase \cite{Pretorius,Campanelli,Baker}. The
ringdown phase is the final stage and describes the evolution of the new
black hole. This new black hole is highly deformed due to the nonlinear
dynamics of the collision. The perturbed black hole emits GWs in the form of
quasinormal radiation and the perturbation theory can be used to calculate
the quasinormal modes (QNMs) \cite{BertiCardosoWill}.

The ringdown phase of a compact binary merger might be better to
study compared to the inspiral or merger phase depending on the
physics being tested \cite{Barack}. The nature of the binary
components shows up only at high post-Newtonian order in the
inspiral phase while investigating the merger phase requires
time-consuming and theory-dependent numerical simulations. On the
other hand, the ringdown phase can be appropriately investigated
by perturbation theory and it is relatively simple to model beyond
Einstein gravity. The perturbation theory of black holes was
started by the pioneering work of Regge and Wheeler \cite{Regge}
and was continued by Zerilli \cite{Zerilli}. They have found the
wave equations of axial and polar perturbations of the
Schwarzschild black hole and examined the dynamical stability of
the black hole under small perturbations. The frequencies of
perturbations are called the QN frequencies (QNFs) and have
been calculated by using analytic and semi-analytic approaches \cite%
{Mashhoon,Schutz,Iyer,KonoplyaWKB}, and also, several numerical methods \cite%
{ChandrasekharDetweiler,Leaver,Horowitz,Naylor,Hatsuda} (see \cite%
{Kokkotas,Berti,Konoplya}\ for reviews on QNMs).

The electromagnetic and gravitational perturbations of the
Schwarzschild black holes have been investigated in \cite{EG}.
Besides, the QNMs for the gravitational and electromagnetic
perturbations in modified gravity are calculated before
\cite{Manfredi}. In the case of conformal gravity, by imposing
perturbations in Minkowskian spacetime, the GWs have been
investigated and the effective energy-momentum tensor of the
gravitational radiation is calculated \cite{RongjiaYang}. The
astrophysical
GWs of inspiralling compact binaries have been investigated \cite%
{Caprini,Holscher,FariaPRD}. Moreover, the scalar, electromagnetic \cite%
{Cosimo}, and axial perturbations \cite{CYChen} of nonsingular
black holes in conformal gravity have been studied. It was shown
that it is possible to find the black hole solutions in Weyl
gravity which are both thermally and dynamically stable under
massive scalar perturbations, and also, the QNMs of this theory of
gravity in asymptotically adS spacetime were obtained
\cite{Hendi}. In addition, the nearly extreme black holes in Weyl
gravity have been studied and an exact formula for the QNFs with
an upper bound on them has been found \cite{Momennia}.

In this paper, our main goal is studying the axial gravitational
perturbations of singular black holes in Weyl gravity in order to
investigate the QN radiation of the ringdown phase. We first obtain a master
wave equation for an arbitrary metric which is conformally related to the
Schwarzschild-(anti) de Sitter (Schwarzschild-(a)dS) spacetime. Then, by
using the Weyl invariance property of the action, and also, the relation
between the Schwarzschild-(a)dS\ black holes and Weyl solutions, we derive
the wave equation of the axial gravitational perturbations of black holes in
Weyl gravity. In addition, we consider both the scalar and electromagnetic
perturbations in the background spacetime of these black holes. Then, we
calculate the QNMs by employing the sixth order WKB approximation and the
asymptotic iteration method (AIM). The time evolution of modes is also
investigated by using the discretization scheme.

\section{4-dimensional black holes in Weyl gravity}

Here, we give a brief review on the four-dimensional black holes in Weyl
gravity and the connection between these solutions and the
Schwarzschild-(a)dS black holes. The action of Weyl gravity is given by \cite%
{PangPope}
\begin{equation}
I=\frac{1}{16\pi }\int d^{4}x\sqrt{-g}C^{\mu \nu \rho \sigma }C_{\mu \nu
\rho \sigma },  \label{I3}
\end{equation}%
where $C_{\mu \nu \rho \sigma }$ is the Weyl conformal tensor with the
following explicit form%
\begin{equation}
C_{\lambda \mu \nu \kappa }=R_{\lambda \mu \nu \kappa }+{\frac{1}{6}}R_{%
\phantom{\alpha}\alpha }^{\alpha }\left[ g_{\lambda \nu }g_{\mu \kappa
}-g_{\lambda \kappa }g_{\mu \nu }\right] -{\frac{1}{2}}\left[ g_{\lambda \nu
}R_{\mu \kappa }-g_{\lambda \kappa }R_{\mu \nu }-g_{\mu \nu }R_{\lambda
\kappa }+g_{\mu \kappa }R_{\lambda \nu }\right] ,
\end{equation}%
and the field equations can be obtained by taking a variation with respect
to the metric tensor $g_{\mu \nu }$%
\begin{equation}
W_{\rho \sigma }=\left( \nabla ^{\mu }\nabla ^{\nu }+\frac{1}{2}R^{\mu \nu
}\right) C_{\rho \mu \nu \sigma }=0,  \label{GFE}
\end{equation}%
which $W_{\mu \nu }$\ is the Bach tensor. It is straightforward to show that
the following $4$-dimensional line element satisfies all components of Eq. (%
\ref{GFE})
\begin{equation}
ds^{2}=-f(r)dt^{2}+f^{-1}(r)dr^{2}+r^{2}\left( d\theta ^{2}+\sin ^{2}\theta
d\varphi ^{2}\right) ,  \label{CS}
\end{equation}%
where the metric function is as follows
\begin{equation}
f\left( r\right) =c+\frac{d}{r}+\frac{c^{2}-1}{3d}r+br^{2},  \label{MF}
\end{equation}%
in which $b$, $c$, and $d$ are three integration constants. It is notable
that in contrast with the Einstein gravity that the cosmological constant
should be considered in the action by hand, it is present as an integration
constant in the Weyl gravity solutions. It is also worthwhile to mention
that one can recover the Schwarzschild-(a)dS black hole by setting $c=1$, $%
d=-2M$, and $b=-\Lambda /3$.

On the other hand, the line element describing the Schwarzschild-(a)dS
spacetime in the radial coordinate $\rho $\ is given by%
\begin{equation}
d\tilde{s}^{2}=-g(\rho )dt^{2}+g^{-1}(\rho )d\rho ^{2}+\rho ^{2}\left(
d\theta ^{2}+\sin ^{2}\theta d\varphi ^{2}\right) ,  \label{SS}
\end{equation}%
in which the metric function is%
\begin{equation}
g\left( \rho \right) =1-\frac{2M}{\rho }-\frac{\Lambda }{3}\rho ^{2},
\label{SMF}
\end{equation}%
where $M$\ denotes the total mass of the black hole and $\Lambda $\ is the
cosmological constant. One can show that there is a conformal relation
between the black hole spacetimes in Weyl gravity (\ref{CS}) and Einstein
theory (\ref{SS}). Indeed, these two spacetimes can be connected to each
other by introducing a conformal factor $S(\rho )$\ so that $ds^{2}=S(\rho )d%
\tilde{s}^{2}$. Every spacetime like Schwarzschild-(a)dS case which is
conformally related to (\ref{CS}) is also a solution of the field equations
of Weyl gravity (\ref{GFE}) since all the metrics that transform conformally
are equivalent. By considering the conformal factor $S(\rho )=\left( 1+q\rho
\right) ^{-2}$ \cite{PangPope}, one can find the relation $\rho =r\left(
1-qr\right) ^{-1}$\ between the radial coordinates $\rho $\ and $r$.
Multiplying (\ref{SS}) by the conformal factor $S(\rho )$\ and replacing $%
\rho $\ by $r$, we can obtain the following relations between the parameters
\cite{PangPope}
\begin{equation}
b=q^{2}\left( 1+2Mq\right) -\frac{\Lambda }{3};\ \ \ c=1+6Mq;\ \ \ d=-2M,\
\label{cond}
\end{equation}%
where $q$\ is an arbitrary constant and we used $\partial _{\rho }=\left(
1-qr\right) ^{2}\partial _{r}$. Therefore, we have a spectrum of conformal
solutions depending on the values of\ $q$. We shall use these relations to
obtain the axial perturbation of Weyl gravity in the coming section.

\section{Gravitational perturbations of Weyl gravity}

Here, we are going to obtain a master equation for the axial gravitational
perturbations of black holes described by the metric function (\ref{MF}).
First, one may note that there is a conformal relation between the line
element (\ref{CS})\ and the Schwarzschild black holes in asymptotically adS
spacetime. Indeed, the spacetime of Weyl solutions and the spacetime of
Schwarzschild-(a)dS solutions are related to each other as $ds^{2}=S(\rho )d%
\tilde{s}^{2}$ so that $ds^{2}$ is the line element of Weyl gravity (\ref{CS}%
), $d\tilde{s}^{2}$ is the line element of the Schwarzschild-(a)dS black
holes (\ref{SS}), and $S(\rho )$ is the conformal factor. Thus, if we obtain
the master equation of black holes conformally related to the
Schwarzschild-(a)dS black holes, it can also describe the gravitational
perturbations of Weyl solutions by replacing the explicit form of $S(\rho )$%
. In other words, if we construct the axial perturbations of $S(\rho )d%
\tilde{s}^{2}$ (for the general form of $S(\rho )$), it is
equivalent to the master equation of Weyl solutions. It is
worthwhile to mention that it is not possible to construct a
second-order master wave equation by
considering small perturbations in (\ref{CS}) and using the field equations (%
\ref{GFE}) because there is fourth-order differential in the field
equations and the linearized field equations are fourth-order.
Here, since our goal is to find a second order wave equation for
Weyl solutions, we shall use the conformal relation between
Schwarzschild-(a)dS black holes and Weyl solutions.

The gravitational perturbations of black holes conformally related to the
Schwarzschild-(a)dS spacetime, i.e. the perturbations of $S(\rho )d\tilde{s}%
^{2}$, is derived in the appendix \ref{appendix}. The master equation of the
axial perturbations of the following spacetime (see APPENDIX \ref{appendix})%
\begin{equation}
ds^{2}=S(\rho )d\tilde{s}^{2}=S\left( \rho \right) \left[ -g\left( \rho
\right) dt^{2}+\frac{d\rho ^{2}}{g\left( \rho \right) }+\rho ^{2}\left(
d\theta ^{2}+\sin ^{2}\theta d\varphi ^{2}\right) \right] ,  \label{Schw1}
\end{equation}%
is given by%
\begin{equation}
\frac{d^{2}\Psi ^{\left( -\right) }\left( \rho _{\ast }\right) }{d\rho
_{\ast }^{2}}+\left[ \omega ^{2}-V_{g}\left( \rho _{\ast }\right) \right]
\Psi ^{\left( -\right) }\left( \rho _{\ast }\right) =0,  \label{we}
\end{equation}%
where $\omega $ is the QN frequency, $\rho _{\ast }=\int \ g^{-1}\left( \rho
\right) d\rho $ is the tortoise coordinate, and the effective potential $%
V_{g}\left( \rho _{\ast }\right) $\ reads%
\begin{equation}
V_{g}\left( \rho _{\ast }\right) =\ g\left( \rho \right) \left[ \frac{\ell
\left( \ell +1\right) }{\rho ^{2}}-\frac{2}{\rho ^{2}}-Z\frac{d}{d\rho }%
\left( \frac{\ g\left( \rho \right) }{Z^{2}}\frac{dZ}{d\rho }\right) \right]
,  \label{Vgr}
\end{equation}%
in which $Z=\rho \sqrt{S\left( \rho \right) }$. Note that the
right-hand side of (\ref{Schw1}) is exactly equal to the Weyl
gravity spacetime (\ref{CS}) under the conditions (\ref{cond})\
whenever we consider the conformal factor $S(\rho )=\left( 1+q\rho
\right) ^{-2}$. As a result, if we consider a coordinate
transformation (in (\ref{Schw1})-(\ref{Vgr})) obeying this
conformal factor, we can obtain the axial perturbations of
singular black holes in Weyl gravity. Thus, we apply the
coordinate transformation $\rho $\ to $r$\ so that $S(\rho
)=\left( 1+q\rho \right) ^{-2}$ and $\rho =r\left( 1-qr\right)
^{-1}$. By considering $\partial _{\rho }=\left( 1-qr\right)
^{2}\partial _{r}$ and $\partial _{r_{\ast }}=g\left( \rho \right)
\left( 1-qr\right) ^{2}\partial _{r}$, one can find the wave
equation (\ref{we})
converts into%
\begin{equation}
\frac{d^{2}\Psi ^{\left( -\right) }\left( r_{\ast }\right) }{dr_{\ast }^{2}}+%
\left[ \omega ^{2}-V_{g}\left( r_{\ast }\right) \right] \Psi ^{\left(
-\right) }\left( r_{\ast }\right) =0,  \label{we2}
\end{equation}%
where $r_{\ast }$\ is the new tortoise coordinate as $dr/dr_{\ast }=\left(
1-qr\right) ^{2}\ \bar{f}\left( r\right) $ with%
\begin{equation}
\bar{f}\left( r\right) =1-\frac{2M\left( 1-qr\right) }{r}-\frac{\Lambda r^{2}%
}{3\left( 1-qr\right) ^{2}}.  \label{Fbar}
\end{equation}

The effective potential (\ref{Vgr}) now is given by%
\begin{equation}
V_{g}\left( r_{\ast }\right) =\ \left( 1-qr\right) ^{2}\bar{f}\left(
r\right) \left[ \frac{\ell \left( \ell +1\right) }{r^{2}}-\frac{2}{r^{2}}-r%
\frac{d}{dr}\left( \frac{\ \bar{f}\left( r\right) }{r^{2}}\left( 1-qr\right)
^{2}\right) \right] ,  \label{Vgr2}
\end{equation}%
which we used $Z=r$ and $\partial _{\rho }=\left( 1-qr\right) ^{2}\partial
_{r}$. Therefore, Eq. (\ref{we2}) is the master equation of the axial
perturbations of \ black holes in the Weyl gravity (\ref{MF}). In addition,
Eq. (\ref{Vgr2}) is the effective potential of perturbations and the free
parameters $b$, $c$, and $d$ of (\ref{MF})\ are related to the parameters of
(\ref{Fbar}) through the conditions (\ref{cond}). It is worthwhile to
mention that for $q=0$, the potential (\ref{Vgr2}) reduces to%
\begin{equation}
V_{g}\left( r_{\ast }\right) =\left( 1-\frac{2M}{r}-\frac{\Lambda r^{2}}{3}%
\right) \left[ \frac{\ell \left( \ell +1\right) }{r^{2}}-\frac{6M}{r^{3}}%
\right] ,  \label{Schwads}
\end{equation}%
which is the effective potential of the axial perturbations of the
Schwarzschild-(a)dS black hole \cite{SchwarzschildadS}, as we expected.

In order to get rid of the free parameter $q$ in (\ref{Vgr2}), one can apply
the conditions (\ref{cond}) in (\ref{Vgr2}) to obtain the following
effective potential
\begin{eqnarray}
V_{g}\left( r_{\ast }\right) &=&\ f\left( r\right) \left[ \frac{\ell \left(
\ell +1\right) }{r^{2}}-\frac{2}{r^{2}}-r\frac{d}{dr}\left( \frac{\ f\left(
r\right) }{r^{2}}\right) \right]  \label{EP} \\
&=&f\left( r\right) \left[ \frac{\ell \left( \ell +1\right) }{r^{2}}+\frac{%
2\left( c-1\right) }{r^{2}}+\frac{c^{2}-1}{3rd}+\frac{3d}{r^{3}}\right] ,
\label{ep}
\end{eqnarray}%
which is a function of the free parameters of conformal gravity ( $r_{\ast
}=\int \ f^{-1}\left( r\right) dr$ being the tortoise coordinate). Now, we
can recover the axial perturbations of the Schwarzschild-(a)dS black hole (%
\ref{Schwads})\ by setting $c=1$, $d=-2M$, and $b=-\Lambda /3$ in (\ref{ep}).

\section{scalar perturbations}

Now, in order to ensure that our calculations of obtaining the axial
perturbations of Weyl gravity are correct, we compare the effective
potential of the massless scalar\ perturbations of black holes in Weyl
gravity obtained by two methods; one is considering the evolution of a
scalar field in the spacetime background (\ref{CS}) directly, and the other
one is multiplying the Schwarzschild spacetime by a conformal factor (\ref%
{Schw1}) and obtaining the related effective potential.

The wave equation and the effective potential of a massless scalar
perturbation in the spacetime background (\ref{CS}) are given by \cite%
{Momennia}
\begin{equation}
\frac{d^{2}\Psi \left( r_{\ast }\right) }{dr_{\ast }^{2}}+\left[ \omega
^{2}-V_{s}\left( r_{\ast }\right) \right] \Psi \left( r_{\ast }\right) =0,
\end{equation}%
\begin{equation}
V_{s}\left( r_{\ast }\right) =f\left( r\right) \left[ \frac{\ell \left( \ell
+1\right) }{r^{2}}+\frac{f^{\prime }\left( r\right) }{r}\right] ,  \label{SP}
\end{equation}%
where $r_{\ast }=\int \ f^{-1}\left( r\right) dr$ is the tortoise
coordinate. On the other hand, the wave equation and the effective potential
of the perturbative conformally related Schwarzschild-(a)dS black holes in $%
\rho $\ coordinate are as follows \cite{Cosimo}%
\begin{equation}
\frac{d^{2}\Psi \left( \rho _{\ast }\right) }{d\rho _{\ast }^{2}}+\left[
\omega ^{2}-V_{s}\left( \rho _{\ast }\right) \right] \Psi \left( \rho _{\ast
}\right) =0,
\end{equation}%
\begin{equation}
V_{s}\left( \rho _{\ast }\right) =g\left( \rho \right) \left[ \frac{\ell
\left( \ell +1\right) }{\rho ^{2}}+\frac{1}{Z}\frac{d}{d\rho }\left( g\left(
\rho \right) \frac{dZ}{d\rho }\right) \right] ,  \label{SPC}
\end{equation}%
where $Z=\rho \sqrt{S\left( \rho \right) }$ and $\rho _{\ast }=\int \
g^{-1}\left( \rho \right) d\rho $ is the tortoise coordinate. Now, we apply
the coordinate transformation $\rho $\ to $r$\ so that $S(\rho )=\left(
1+q\rho \right) ^{-2}$ and $\rho =r\left( 1-qr\right) ^{-1}$, as was
described. Thus, the effective potential (\ref{SPC})\ reduces to%
\begin{equation}
V_{s}\left( r_{\ast }\right) =\ \left( 1-qr\right) ^{2}\bar{f}\left(
r\right) \left[ \frac{\ell \left( \ell +1\right) }{r^{2}}+\frac{1}{r}\frac{d%
}{dr}\left( \bar{f}\left( r\right) \left( 1-qr\right) ^{2}\right) \right] .
\label{SP1}
\end{equation}

It is straightforward to show that $\left( 1-qr\right) ^{2}\bar{f}\left(
r\right) $\ is equal to the metric function of Weyl gravity (\ref{MF}) with
the help of obtained conditions (\ref{cond}), and thus the effective
potential (\ref{SP1}) is equal to Eq. (\ref{SP}). Therefore, this comparison
of scalar perturbations shows that our calculations of obtaining the axial
perturbations of Weyl solutions given in the previous section are indeed
correct.

\section{electromagnetic perturbations}

The wave equation and effective potential of electromagnetic perturbation in
the spacetime background (\ref{CS}) are given by (see APPENDIX \ref{app})%
\begin{equation}
\frac{d^{2}\Psi \left( r_{\ast }\right) }{dr_{\ast }^{2}}+\left[ \omega
^{2}-V_{e}\left( r_{\ast }\right) \right] \Psi \left( r_{\ast }\right) =0,
\end{equation}%
\begin{equation}
V_{e}\left( r_{\ast }\right) =f\left( r\right) \frac{\ell \left( \ell
+1\right) }{r^{2}},  \label{Ve}
\end{equation}%
where $r_{\ast }=\int \ f^{-1}\left( r\right) dr$ is the tortoise
coordinate. On the other hand, the master equation of the perturbative
conformally related Schwarzschild-(a)dS black holes in $\rho $\ coordinate
is \cite{Cosimo}%
\begin{equation}
\frac{d^{2}\Psi \left( \rho _{\ast }\right) }{d\rho _{\ast }^{2}}+\left[
\omega ^{2}-V_{e}\left( \rho _{\ast }\right) \right] \Psi \left( \rho _{\ast
}\right) =0,  \label{WE}
\end{equation}%
with%
\begin{equation}
V_{e}\left( \rho _{\ast }\right) =g\left( \rho \right) \frac{\ell \left(
\ell +1\right) }{\rho ^{2}},  \label{Ver}
\end{equation}%
where $\rho _{\ast }=\int \ g^{-1}\left( \rho \right) d\rho $ is the
tortoise coordinate. By applying the coordinate transformation $\rho $\ to $%
r $\ such that $S(\rho )=\left( 1+q\rho \right) ^{-2}$ and $\rho =r\left(
1-qr\right) ^{-1}$, the effective potential (\ref{Ver})\ reduces to%
\begin{equation}
V_{e}\left( r_{\ast }\right) =\ \left( 1-qr\right) ^{2}\bar{f}\left(
r\right) \frac{\ell \left( \ell +1\right) }{r^{2}},  \label{nVer}
\end{equation}%
which $\left( 1-qr\right) ^{2}\bar{f}\left( r\right) $\ is equal to the
metric function of Weyl gravity (\ref{MF}) by considering the conditions (%
\ref{cond}). Therefore, the effective potentials (\ref{Ve}) and (\ref{nVer}%
)\ are the same.\

It is worthwhile to mention that scalar, electromagnetic, and gravitational
perturbations of Weyl gravity can be collected and described by the
following master equation%
\begin{equation}
\frac{d^{2}\Psi \left( r_{\ast }\right) }{dr_{\ast }^{2}}+\left[ \omega
^{2}-V_{\ell }\left( r_{\ast }\right) \right] \Psi \left( r_{\ast }\right)
=0,  \label{Master}
\end{equation}%
with the potential%
\begin{equation}
V_{\ell }\left( r_{\ast }\right) =f\left( r\right) \left[ \frac{\ell \left(
\ell +1\right) }{r^{2}}+\left( 1-s^{2}\right) \left( \frac{\left(
4-s^{2}\right) b}{2}-\frac{d}{r^{3}}-\frac{s\left( c-1\right) }{3r^{2}}%
\right) +\left( 1-s\right) ^{2}\frac{c^{2}-1}{3rd}\right] ,  \label{V}
\end{equation}%
where $s=0,1,2$ is the spin of the perturbing field and we used
the effective potentials given in Eqs. (\ref{ep}), (\ref{SP}), and
(\ref{Ve}) to
obtain this equation. By inserting $c=1$, $d=-2M$, and $b=-\Lambda /3$ into (%
\ref{V}), one can obtain
\begin{equation}
V_{\ell }\left( r_{\ast }\right) =f\left( r\right) \left[ \frac{\ell \left(
\ell +1\right) }{r^{2}}+\left( 1-s^{2}\right) \left( \frac{2M}{r^{3}}-\frac{%
\left( 4-s^{2}\right) \Lambda }{6}\right) \right] ,
\end{equation}%
for the Schwarzschild-(a)dS black holes \cite{Berti}.

\section{Quasinormal modes}

Here, we consider the master equation (\ref{Master})\ with the effective
potential (\ref{V})\ for $s=0,1,2$\ as the results of perturbations of Weyl
gravity. The spectrum of QNMs is the solution of the wave equation (\ref%
{Master}) and this spectrum becomes discrete when we impose some proper
boundary conditions. The boundary conditions are applied to the modes $\Psi
\left( r_{\ast }\right) $\ at $r_{\ast }\rightarrow \pm \infty $\ which can
be obtained by studying the flux of radiation detected by physical observers
near the event horizon and cosmological horizon. The observers detect
outgoing waves at the cosmological horizon and incoming radiation at the
event horizon%
\begin{equation}
\left\{
\begin{array}{cc}
\Psi \left( r_{\ast }\right) \sim e^{-i\omega r_{\ast }}, & r_{\ast
}\rightarrow -\infty \ \left( r\rightarrow r_{e}\right) \\
\Psi \left( r_{\ast }\right) \sim e^{i\omega r_{\ast }}, & r_{\ast
}\rightarrow +\infty \ \left( r\rightarrow r_{c}\right)%
\end{array}%
\right. ,
\end{equation}%
where $r_{e}$ is the event horizon and $r_{c}$ denotes the cosmological
horizon.

Now, we concentrate our attention on the conformal-dS solutions ($b<0$) and
calculate the QN frequencies by using the WKB formula as a semi-analytic
approach and AIM as a numerical method. The WKB approximation is based on
the matching of WKB expansion of the modes $\Psi \left( r_{\ast }\right) $
at the event horizon and cosmological horizon with the Taylor expansion near
the peak of the potential barrier. This method can be used for an effective
potential that forms a potential barrier with a single peak. It was first
applied to the problem of scattering around black holes \cite{Schutz}, and
then extended to the $3$rd order \cite{Iyer}, $6$th\ order \cite{KonoplyaWKB}%
, and $13$th\ order \cite{Matyjasek}. The WKB formula is given by%
\begin{equation}
\omega ^{2}=V_{0}+\sum_{j=1}\Omega _{2j}-i\left( n+\frac{1}{2}\right) \sqrt{%
-2V_{0}^{\prime \prime }}\left( 1+\sum_{j=1}\Omega _{2j+1}\right) ,
\label{WKB}
\end{equation}%
where $n$ is the overtone number, $V_{0}$\ is the value of the effective
potential at its local maximum, and $\Omega _{k}$'s denote the $k$th order
of the approximation and depend on the value of the effective potential and
its derivatives at the local maximum. It is notable that the explicit form
of the WKB corrections is given in \cite{Iyer,KonoplyaWKB}. We shall use
this formula up to the sixth order as a semi-analytical approach to obtain
the QNFs of perturbations.

On the other hand, the AIM has been employed for solving the eigenvalue
problems and second-order differential equations \cite{Ciftci,CiftciHall}.
It was also shown that the improved AIM can be used as an accurate numerical
method for calculating the QNMs \cite{Naylor,ChoNaylor}. Here, we will use
this method up to $15$\ iterations as a numerical approach to obtain the
QNFs of perturbations.

In addition, we can investigate the contribution of all modes by using the
time-domain integration of the wavelike equation (\ref{Master}). The
time-domain profile of modes shows the time evolution of modes at the
ringdown stage and the behavior of the asymptotic tails at late times. In
order to obtain the time evolution of modes, we follow the discrimination
scheme given in \cite{Pullin}. The perturbation equation (\ref{Master})
takes the following form%
\begin{equation}
-4\frac{\partial ^{2}\Psi \left( u,v\right) }{\partial u\partial v}=V_{\ell
}\left( u,v\right) \Psi \left( u,v\right) ,
\end{equation}%
in terms of the light-cone coordinates $u=t-x$\ and $v=t+x$, and $\Psi $\
assumed to have time dependence $e^{-i\omega t}$. We can obtain the
time-domain profile of modes by integrating this equation on the small grids
from the two null surfaces $u=u_{0}$\ and $v=v_{0}$. One can obtain the
evolution equation in the light-cone coordinates by applying the time
evolution operator on $\Psi \left( u,v\right) $\ and expanding this operator
for sufficiently small grids
\begin{eqnarray}
\Psi \left( u+\Delta ,v+\Delta \right) &=&\Psi \left( u+\Delta ,v\right)
+\Psi \left( u,v+\Delta \right) -\Psi \left( u,v\right)  \notag \\
&&-\frac{\Delta ^{2}}{8}\left[ V_{\ell }\left( u+\Delta ,v\right) \Psi
\left( u+\Delta ,v\right) +V_{\ell }\left( u,v+\Delta \right) \Psi \left(
u,v+\Delta \right) \right] ,
\end{eqnarray}%
which $\Delta $\ is the step size of the grids. We shall obtain the time
evolution of perturbations with a Gaussian wave packet on the surfaces $%
u=u_{0}$\ and $v=v_{0}$ as initial data.

The QNFs of scalar ($s=0$), electromagnetic ($s=1$), and gravitational ($s=2$%
) perturbations are presented in tables $I-IV$. We calculated the lowest
frequencies (tables $I$ and $II$) and the second overtone (tables $III$ and $%
IV$)\ for some values of the free parameters $b$, $c$, $d$, and $\ell $. The
frequencies are written as $\omega =\omega _{r}-i\omega _{i}$ where $\omega
_{r}$ ($\omega _{i}$) is the real (imaginary) part of the frequencies. The
obtained frequencies for gravitational perturbation show that the WKB
formula and the AI method are in a good agreement. The modes of
gravitational perturbation live longer with lower frequency compared to the
scalar and electromagnetic perturbations. In addition, the all kinds of
perturbations decay faster with more oscillations by increasing $d$\ and/or $%
b$. However, the effective potential is positive and all frequencies have a
negative imaginary part which indicates that these kinds of perturbations
will decay with time, and thus, the spacetime is stable.

\begin{center}
\begin{tabular}{|c|c|c|c|c|c|c|}
\hline\hline
$b$ & $c$ & $d$ &  & $s=2$ & $s=1$ & $s=0$ \\ \hline\hline
$-0.10$ & $0.30$ & $-1.0$ &  & $%
\begin{array}{c}
0.155204-0.0397912i \\
0.155204-0.0397909i%
\end{array}%
$ & $%
\begin{array}{c}
0.191879-0.0399914i \\
0.191879-0.0399918i%
\end{array}%
$ & $%
\begin{array}{c}
0.192944-0.0404160i \\
0.192944-0.0404143i%
\end{array}%
$ \\ \hline
$-0.10$ & $0.30$ & $-0.9$ &  & $%
\begin{array}{c}
0.344680-0.0863702i \\
0.344684-0.0863653i%
\end{array}%
$ & $%
\begin{array}{c}
0.424112-0.0882848i \\
0.424111-0.0882865i%
\end{array}%
$ & $%
\begin{array}{c}
0.433858-0.0911394i \\
0.433857-0.0911551i%
\end{array}%
$ \\ \hline
$-0.10$ & $0.30$ & $-0.8$ &  & $%
\begin{array}{c}
0.502599-0.123024i \\
0.502617-0.122989i%
\end{array}%
$ & $%
\begin{array}{c}
0.615836-0.128076i \\
0.615836-0.128079i%
\end{array}%
$ & $%
\begin{array}{c}
0.639933-0.133019i \\
0.639943-0.133023i%
\end{array}%
$ \\ \hline
$-0.10$ & $0.35$ & $-1.0$ &  & $%
\begin{array}{c}
0.206948-0.0528026i \\
0.206948-0.0528016i%
\end{array}%
$ & $%
\begin{array}{c}
0.255592-0.0532524i \\
0.255592-0.0532532i%
\end{array}%
$ & $%
\begin{array}{c}
0.257955-0.0541371i \\
0.257953-0.0541365i%
\end{array}%
$ \\ \hline
$-0.10$ & $0.40$ & $-1.0$ &  & $%
\begin{array}{c}
0.246709-0.0626906i \\
0.246710-0.0626885i%
\end{array}%
$ & $%
\begin{array}{c}
0.304454-0.0634129i \\
0.304454-0.0634139i%
\end{array}%
$ & $%
\begin{array}{c}
0.308203-0.0647413i \\
0.308199-0.064744i%
\end{array}%
$ \\ \hline
$-0.09$ & $0.30$ & $-1.0$ &  & $%
\begin{array}{c}
0.248847-0.0630602i \\
0.248849-0.0630587i%
\end{array}%
$ & $%
\begin{array}{c}
0.306884-0.0639179i \\
0.306884-0.0639190i%
\end{array}%
$ & $%
\begin{array}{c}
0.311359-0.0654485i \\
0.311355-0.0654544i%
\end{array}%
$ \\ \hline
$-0.08$ & $0.30$ & $-1.0$ &  & $%
\begin{array}{c}
0.316319-0.0791611i \\
0.316323-0.0791561i%
\end{array}%
$ & $%
\begin{array}{c}
0.389118-0.0809956i \\
0.389118-0.0809973i%
\end{array}%
$ & $%
\begin{array}{c}
0.398428-0.0836725i \\
0.398427-0.0836869i%
\end{array}%
$ \\ \hline\hline
\end{tabular}%
\vspace{0.2cm}

Table $I$: The fundamental QNMs for $\ell =2$, calculated by using the AIM
method (first row) and WKB formula (second row).

\begin{tabular}{|c|c|c|c|c|c|c|}
\hline\hline
$b$ & $c$ & \multicolumn{1}{|c|}{$d$} &  & $s=2$ & $s=1$ & $s=0$ \\
\hline\hline
$-0.10$ & $0.30$ & \multicolumn{1}{|c|}{$-1.0$} &  & $%
\begin{array}{c}
0.249970-0.0398942i \\
0.249970-0.0398943i%
\end{array}%
$ & \multicolumn{1}{|c|}{$%
\begin{array}{c}
0.274336-0.0400070i \\
0.274336-0.0400071i%
\end{array}%
$} & $%
\begin{array}{c}
0.275107-0.0402185i \\
0.275107-0.0402184i%
\end{array}%
$ \\ \hline
$-0.10$ & $0.30$ & \multicolumn{1}{|c|}{$-0.9$} &  & $%
\begin{array}{c}
0.553216-0.0873844i \\
0.553216-0.0873848i%
\end{array}%
$ & \multicolumn{1}{|c|}{$%
\begin{array}{c}
0.606668-0.0884178i \\
0.606668-0.0884180i%
\end{array}%
$} & $%
\begin{array}{c}
0.613560,-0.0898396i \\
0.613569-0.0898409i%
\end{array}%
$ \\ \hline
$-0.10$ & $0.30$ & \multicolumn{1}{|c|}{$-0.8$} &  & $%
\begin{array}{c}
0.804179-0.125785i \\
0.804179-0.125785i%
\end{array}%
$ & \multicolumn{1}{|c|}{$%
\begin{array}{c}
0.881325-0.128389i \\
0.881325-0.1283893i%
\end{array}%
$} & $%
\begin{array}{c}
0.898233-0.130873i \\
0.898234-0.130874i%
\end{array}%
$ \\ \hline
$-0.10$ & $0.35$ & \multicolumn{1}{|c|}{$-1.0$} &  & $%
\begin{array}{c}
0.333071-0.0530388i \\
0.333071-0.0530390i%
\end{array}%
$ & \multicolumn{1}{|c|}{$%
\begin{array}{c}
0.365476-0.0532890i \\
0.365476-0.0532891i%
\end{array}%
$} & $%
\begin{array}{c}
0.367175-0.0537292i \\
0.367175-0.0537291i%
\end{array}%
$ \\ \hline
$-0.10$ & $0.40$ & \multicolumn{1}{|c|}{$-1.0$} &  & $%
\begin{array}{c}
0.396847-0.0630772i \\
0.396847-0.0630774i%
\end{array}%
$ & \multicolumn{1}{|c|}{$%
\begin{array}{c}
0.435398-0.0634744i \\
0.435398-0.0634745i%
\end{array}%
$} & $%
\begin{array}{c}
0.438083-0.0641348i \\
0.438083-0.0641349i%
\end{array}%
$ \\ \hline
$-0.09$ & $0.30$ & \multicolumn{1}{|c|}{$-1.0$} &  & $%
\begin{array}{c}
0.400063-0.0635081i \\
0.400063-0.0635083i%
\end{array}%
$ & \multicolumn{1}{|c|}{$%
\begin{array}{c}
0.438876-0.0639809i \\
0.438876-0.0639810i%
\end{array}%
$} & $%
\begin{array}{c}
0.442073-0.0647420i \\
0.442072-0.0647424i%
\end{array}%
$ \\ \hline
$-0.08$ & $0.30$ & \multicolumn{1}{|c|}{$-1.0$} &  & $%
\begin{array}{c}
0.507603-0.0801345i \\
0.507604-0.0801349i%
\end{array}%
$ & \multicolumn{1}{|c|}{$%
\begin{array}{c}
0.556626-0.0811223i \\
0.556626-0.0811225i%
\end{array}%
$} & $%
\begin{array}{c}
0.563213-0.0824560i \\
0.563213-0.0824572i%
\end{array}%
$ \\ \hline\hline
\end{tabular}%
\vspace{0.2cm}

Table $II$: The fundamental QNMs for $\ell =3$, calculated by using the AIM
method (first row) and WKB formula (second row).

\begin{tabular}{|c|c|c|c|c|c|c|}
\hline\hline
$b$ & $c$ & \multicolumn{1}{|c|}{$d$} &  & $s=2$ & $s=1$ & $s=0$ \\
\hline\hline
$-0.10$ & $0.30$ & \multicolumn{1}{|c|}{$-1.0$} &  & $%
\begin{array}{c}
0.154934-0.119369i \\
0.154934-0.119368i%
\end{array}%
$ & \multicolumn{1}{|c|}{$%
\begin{array}{c}
0.191502-0.119978i \\
0.191503-0.119979i%
\end{array}%
$} & $%
\begin{array}{c}
0.192318-0.121313i \\
0.192313-0.121311i%
\end{array}%
$ \\ \hline
$-0.10$ & $0.30$ & \multicolumn{1}{|c|}{$-0.9$} &  & $%
\begin{array}{c}
0.342039-0.258939i \\
0.342060-0.258873i%
\end{array}%
$ & \multicolumn{1}{|c|}{$%
\begin{array}{c}
0.420737-0.264990i \\
0.420737-0.264993i%
\end{array}%
$} & $%
\begin{array}{c}
0.431318-0.274284i \\
0.431374-0.274289i%
\end{array}%
$ \\ \hline
$-0.10$ & $0.30$ & \multicolumn{1}{|c|}{$-0.8$} &  & $%
\begin{array}{c}
0.495174-0.368250i \\
0.495149-0.367803i%
\end{array}%
$ & \multicolumn{1}{|c|}{$%
\begin{array}{c}
0.607537-0.384809i \\
0.607535-0.384817i%
\end{array}%
$} & $%
\begin{array}{c}
0.636672-0.399495i \\
0.636579-0.399406i%
\end{array}%
$ \\ \hline
$-0.10$ & $0.35$ & \multicolumn{1}{|c|}{$-1.0$} &  & $%
\begin{array}{c}
0.206263-0.158394i \\
0.206265-0.158390i%
\end{array}%
$ & \multicolumn{1}{|c|}{$%
\begin{array}{c}
0.254695-0.159773i \\
0.254696-0.159774i%
\end{array}%
$} & $%
\begin{array}{c}
0.256703-0.162615i \\
0.256695-0.162638i%
\end{array}%
$ \\ \hline
$-0.10$ & $0.40$ & \multicolumn{1}{|c|}{$-1.0$} &  & $%
\begin{array}{c}
0.245478-0.188046i \\
0.245482-0.188033i%
\end{array}%
$ & \multicolumn{1}{|c|}{$%
\begin{array}{c}
0.302929-0.190277i \\
0.302930-0.190279i%
\end{array}%
$} & $%
\begin{array}{c}
0.306359-0.194596i \\
0.306362-0.194640i%
\end{array}%
$ \\ \hline
$-0.09$ & $0.30$ & \multicolumn{1}{|c|}{$-1.0$} &  & $%
\begin{array}{c}
0.247683-0.189134i \\
0.247689-0.189124i%
\end{array}%
$ & \multicolumn{1}{|c|}{$%
\begin{array}{c}
0.305322-0.191793i \\
0.305322-0.191795i%
\end{array}%
$} & $%
\begin{array}{c}
0.309593-0.196795i \\
0.309608-0.196847i%
\end{array}%
$ \\ \hline
$-0.08$ & $0.30$ & \multicolumn{1}{|c|}{$-1.0$} &  & $%
\begin{array}{c}
0.313782-0.237311i \\
0.313803-0.237242i%
\end{array}%
$ & \multicolumn{1}{|c|}{$%
\begin{array}{c}
0.385897-0.243122i \\
0.385897-0.243124i%
\end{array}%
$} & $%
\begin{array}{c}
0.396113-0.251816i \\
0.396163-0.251809i%
\end{array}%
$ \\ \hline\hline
\end{tabular}%
\vspace{0.2cm}

Table $III$: The QNMs for $n=1$\ and $\ell =2$, calculated by using the AIM
method (first row) and WKB formula (second row).

\begin{tabular}{|c|c|c|c|c|c|c|}
\hline\hline
$b$ & $c$ & $d$ &  & $s=2$ & $s=1$ & $s=0$ \\ \hline\hline
$-0.10$ & $0.30$ & $-1.0$ &  & $%
\begin{array}{c}
0.249738-0.119682i \\
0.249738-0.119682i%
\end{array}%
$ & \multicolumn{1}{|c|}{$%
\begin{array}{c}
0.274064-0.120023i \\
0.274064-0.120023i%
\end{array}%
$} & $%
\begin{array}{c}
0.274748-0.120672i \\
0.274747-0.120672i%
\end{array}%
$ \\ \hline
$-0.10$ & $0.30$ & $-0.9$ &  & $%
\begin{array}{c}
0.550996-0.262139i \\
0.550997-0.262138i%
\end{array}%
$ & \multicolumn{1}{|c|}{$%
\begin{array}{c}
0.604256-0.265327i \\
0.604256-0.265327i%
\end{array}%
$} & $%
\begin{array}{c}
0.611367-0.269787i \\
0.611371-0.269791i%
\end{array}%
$ \\ \hline
$-0.10$ & $0.30$ & $-0.8$ &  & $%
\begin{array}{c}
0.798271-0.377357i \\
0.798268-0.377350i%
\end{array}%
$ & \multicolumn{1}{|c|}{$%
\begin{array}{c}
0.875436-0.385472i \\
0.875436-0.385473i%
\end{array}%
$} & $%
\begin{array}{c}
0.893985-0.392979i \\
0.893982-0.392971i%
\end{array}%
$ \\ \hline
$-0.10$ & $0.35$ & $-1.0$ &  & $%
\begin{array}{c}
0.332501-0.159116i \\
0.332501-0.159116i%
\end{array}%
$ & \multicolumn{1}{|c|}{$%
\begin{array}{c}
0.364830-0.159876i \\
0.364830-0.159876i%
\end{array}%
$} & $%
\begin{array}{c}
0.366401-0.161240i \\
0.366400-0.161242i%
\end{array}%
$ \\ \hline
$-0.10$ & $0.40$ & $-1.0$ &  & $%
\begin{array}{c}
0.395853-0.189233i \\
0.395853-0.189234i%
\end{array}%
$ & \multicolumn{1}{|c|}{$%
\begin{array}{c}
0.434304-0.190444i \\
0.434304-0.190444i%
\end{array}%
$} & $%
\begin{array}{c}
0.436862-0.192505i \\
0.436862-0.192508i%
\end{array}%
$ \\ \hline
$-0.09$ & $0.30$ & $-1.0$ &  & $%
\begin{array}{c}
0.399070-0.190519i \\
0.399071-0.190519i%
\end{array}%
$ & \multicolumn{1}{|c|}{$%
\begin{array}{c}
0.437755-0.191964i \\
0.437755-0.191965i%
\end{array}%
$} & $%
\begin{array}{c}
0.440857-0.194347i \\
0.440857-0.194351i%
\end{array}%
$ \\ \hline
$-0.08$ & $0.30$ & $-1.0$ &  & $%
\begin{array}{c}
0.505476-0.240390i \\
0.505477-0.240388i%
\end{array}%
$ & \multicolumn{1}{|c|}{$%
\begin{array}{c}
0.554326-0.243440i \\
0.554326-0.243440i%
\end{array}%
$} & $%
\begin{array}{c}
0.561149-0.247621i \\
0.561153-0.247624i%
\end{array}%
$ \\ \hline\hline
\end{tabular}%
\vspace{0.2cm}

Table $IV$: The QNMs for $n=1$\ and $\ell =3$, calculated by using the AIM
method (first row) and WKB formula (second row).
\end{center}

\begin{figure}[tbp]
$%
\begin{array}{ccc}
\epsfxsize=7.5cm \epsffile{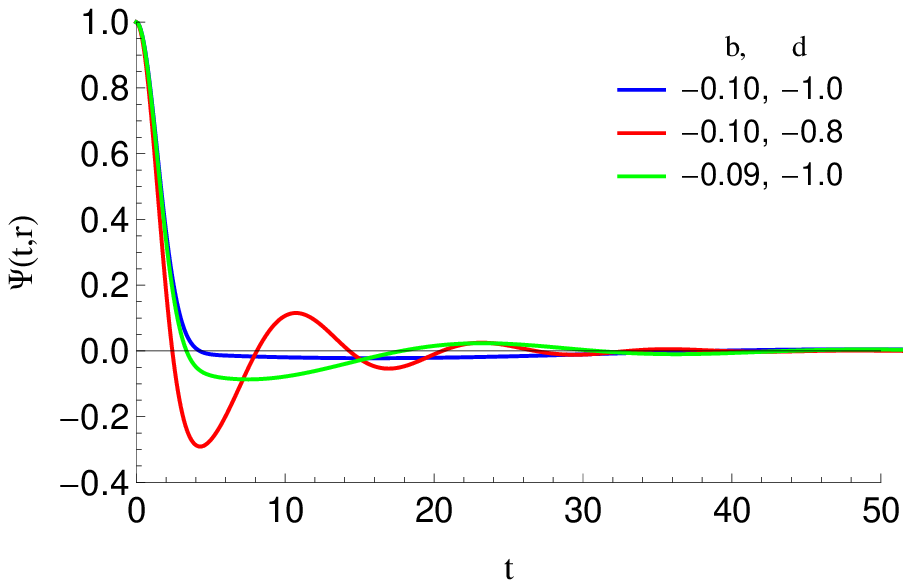} & \epsfxsize=7.5cm \epsffile{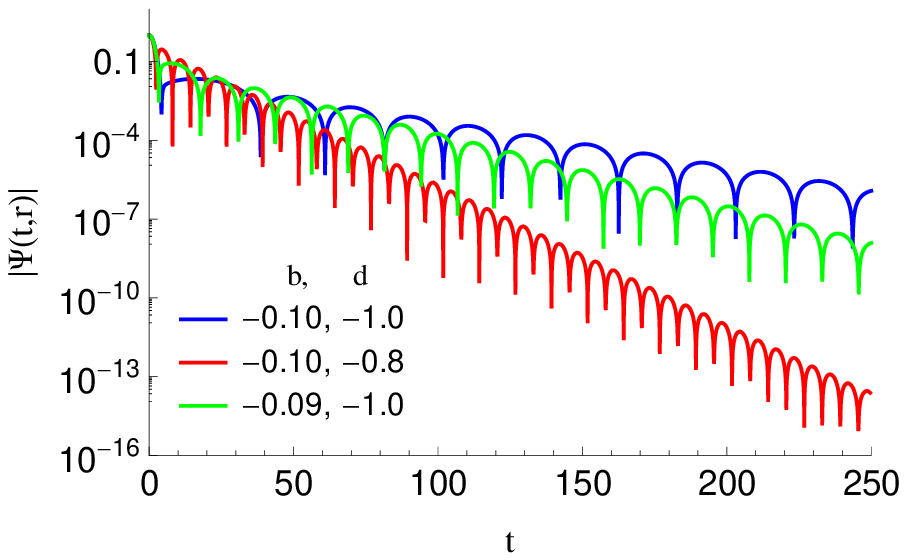} &
\end{array}
$%
\caption{This figure evaluated at $r=2$ ($r_{e}<2<r_{c}$) for $\ell =2$ and $%
c=0.3$. The left (right) panel indicates the (absolute) value of the wave
function $\Psi \left( t,r\right) $ of gravitational perturbations as a
function of time}
\label{Fig}
\end{figure}

\begin{figure}[tbp]
$%
\begin{array}{ccc}
\epsfxsize=7.5cm \epsffile{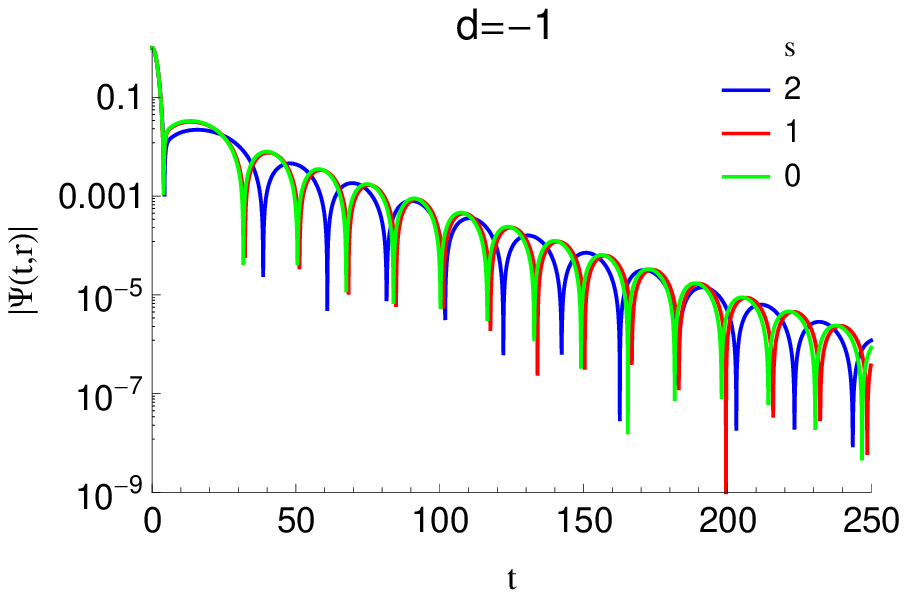} & \epsfxsize=7.5cm \epsffile{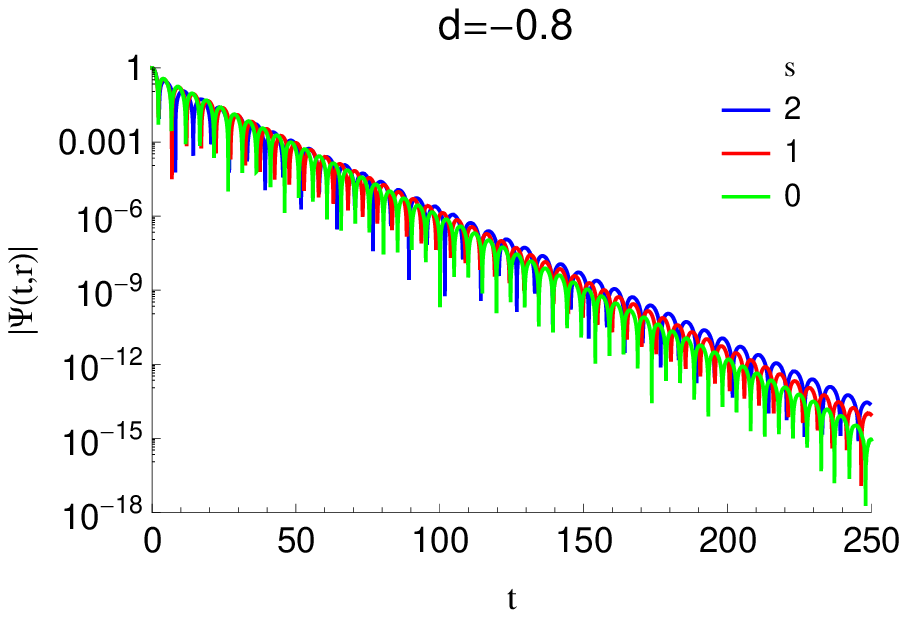}
&
\end{array}
$%
\caption{This figure evaluated at $r=2$ for $\ell =2$, $b=-0.1$, and $c=0.3$%
. It shows the QN modes of scalar ($s=0$) and electromagnetic ($s=1$)
perturbations alongside the gravitational ones ($s=2$). By comparing the
left panel and right panel we find that the modes of all perturbations decay
faster with more oscillations by increasing $d$}
\label{GESfig}
\end{figure}

The time-domain profile of modes for different perturbations is presented in
Figs. \ref{Fig}\ and \ref{GESfig} for some fixed values of the free
parameters. According to the Fig. \ref{Fig}, we can observe that the QN
oscillations of the wave function $\Psi \left( t,r\right) $\ at the ringdown
phase of gravitational perturbations for early and intermediate times. This
figure confirms that the wave function oscillates with a frequency that
increases and decay faster with increasing $b$ and/or $d$. In addition, the
time evolution of modes for scalar and electromagnetic perturbations is
illustrated in Fig. \ref{GESfig}. This figure shows that the QN oscillations
of gravitational perturbation live longer with lower frequency compared to
the scalar and electromagnetic perturbations.

\begin{figure}[tbp]
$%
\begin{array}{ccc}
\epsfxsize=7.5cm \epsffile{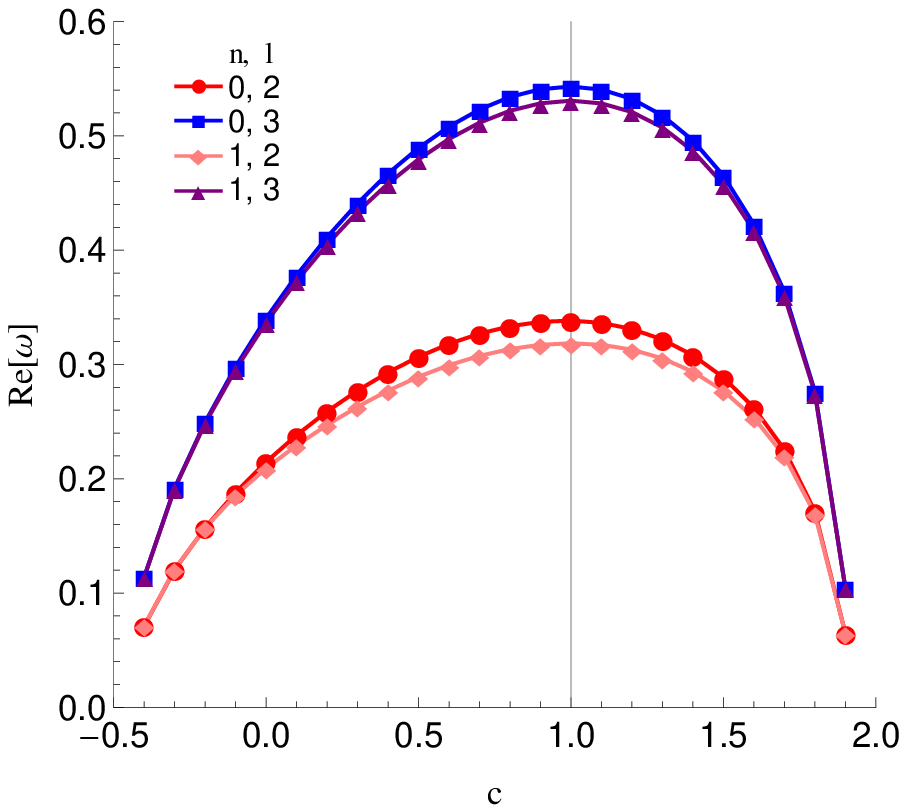} & \epsfxsize=7.5cm \epsffile{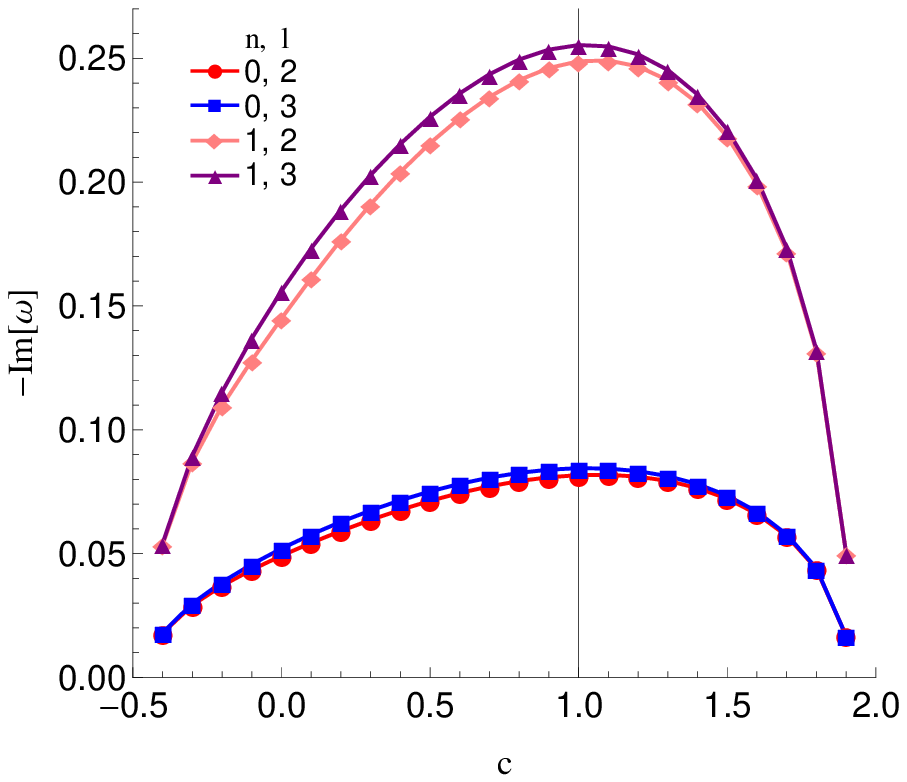}
&
\end{array}
$%
\caption{$Re[\protect\omega ]$ and $Im[\protect\omega ]$ as a function of $c$
for $\Lambda =0.02$ and unit mass calculated by using the sixth order WKB
formula. The vertical line indicates the frequencies of the Schwarzschild-dS
black hole. Both the real and imaginary parts of QN frequencies decrease
whenever $c$ deviates from one}
\label{Cfig}
\end{figure}
In order to obtain the deviations of Weyl solutions from the
Schwarzschild-dS black holes, we compare the QNM spectra of the both cases.
To do so, we first consider $d=-2M$\ and $b=-\Lambda /3$ in the metric
function (\ref{MF}), and then calculate the gravitational QN modes. In this
way, the metric function of conformal gravity takes the form
\begin{equation}
f\left( r\right) =c-\frac{2M}{r}-\frac{c^{2}-1}{6M}r-\frac{\Lambda }{3}r^{2},
\end{equation}%
and thus, the value of $c$\ characterizes the deviation. As $c$\ gets away
from $1$, both the real and imaginary parts of frequencies decrease which
shows that the perturbations in the conformal black holes' background live
longer compared to the Schwarzschild ones (see Fig. \ref{Cfig}). It is
notable that in order to have black hole solutions, the value of $c$\ cannot
be chosen from $c\leq -0.5$\ and $c\geq 2$, and also, the allowed range for
nearly extreme black holes is $-1<c<2$\ \cite{Momennia}.

In addition, the exact relation for QNFs in the eikonal limit ($\ell
\rightarrow \infty $) can be obtained by the first order WKB formula (\ref%
{WKB})%
\begin{equation}
\omega ^{2}=V_{0}-i\left( n+\frac{1}{2}\right) \sqrt{-2V_{0}^{\prime \prime }%
},
\end{equation}%
and in this regime, the effective potential (\ref{V}) is given by%
\begin{equation}
V_{\ell }\left( r_{\ast }\right) \sim f\left( r\right) \frac{\ell ^{2}}{r^{2}%
},
\end{equation}%
which is still a function of $c$ due to the presence of $f\left( r\right) $.
Interestingly, we find that these black holes, unlike the non-singular black
holes in conformal gravity \cite{CYChen}, can be distinguished from the
Schwarzschild ones even in the eikonal limit.

\subsection{Error estimation of QN frequencies}

Although the WKB formula gives the best accuracy at $\ell >n$ and
gives us an accurate and economic way to compute the QN
frequencies, this method does not always give a reliable result
and neither guarantees a good estimation for the error
\cite{KonoplyaWKB,Zhidenko}. In addition, we cannot
always~increase the WKB order to obtain a better approximation for
the frequency. In practice, there is some order that the WKB
formula (\ref{WKB}) provides the best approximation, and the error
increases\ as the order of the formula increases. On the other
hand, since the AIM relies upon a specialized barrier function,
there is no systematic way to estimate the errors or to improve
the accuracy \cite{Naylor}. In order to estimate the error of the
WKB approximation, we can compare two sequential orders of the
formula (\ref{WKB}). However, since each WKB order correction
affects either real or imaginary part of the squared frequencies,
we should use the following quantity%
\begin{equation}
\Delta _{k}=\frac{\left\vert \omega _{k+1}-\omega _{k-1}\right\vert }{2},
\label{error}
\end{equation}%
for the error estimation of $\omega _{k}$ that is calculated with
the WKB formula of the order $k$. This quantity allows determining
the WKB order in which the error is minimal. In the case of the
Schwarzschild black holes, it is shown that the error estimation
(\ref{error})\ provides a very good
estimation of the error order, satisfying \cite{Zhidenko}%
\begin{equation}
\Delta _{k}\gtrsim \left\vert \omega -\omega _{k}\right\vert ,  \label{abs}
\end{equation}%
where $\omega $\ is an accurate value of the QN frequency calculated by
using numerical methods. Thus, we can use the error estimation (\ref{error})
to find the order of the absolute error and determine the order which gives
the most accurate approximation for the QN mode.

In the tables $V$ and $VI$, we summarize the error estimation of the WKB
formula for the fundamental and first overtone QNMs of fixed $\ell =2$, $%
b=-0.08$, $c=0.3$, and $d=-1$ (related to the last line of tables $I$ and $%
III$). From table $V$, we can see that the best order of the WKB
formula for calculating the QN frequency of gravitational
perturbations is $7$th-order whereas the QN frequency of
electromagnetic perturbations has the best accuracy with the help
of $12$th-order. In the case of scalar perturbations, the error of
the WKB formula of $10$th-order is minimal. Table $VI$\ shows the
same results for the first overtone frequency, except for
gravitational perturbations. The QN frequency of gravitational
perturbations has the best
accuracy with the help of $8$th-order. If we assume that the band (\ref{abs}%
) is also correct for the Weyl solutions, since the maximum estimation of
the error for the $6$th-order WKB formula is of order $10^{-6}$ and $10^{-5}$%
, up to $4$ ($3$) digits of the frequency in the last line of table $I$ ($III
$) are reliable. Our calculations based on (\ref{abs}), not shown here due
to economic reason, show that the frequencies given in tables $I-IV$ are
reliable up to a minimum of $3$ digits.

\begin{center}
\begin{tabular}{|c|c|c|c|c|c|c|}
\hline\hline
$k$ & $\omega _{k}$ $\left( s=2\right) $ & $\Delta _{k}$ $\left( s=2\right) $
& $\omega _{k}$ $\left( s=1\right) $ & $\Delta _{k}$ $\left( s=1\right) $ & $%
\omega _{k}$ $\left( s=0\right) $ & $\Delta _{k}$ $\left( s=0\right) $ \\
\hline\hline
$2$ & $0.316898-0.0817235i$ & $9.6577\times 10^{-3}$ & $0.389434-0.0830497i$
& $8.4971\times 10^{-3}$ & $0.398835-0.0856398i$ & $8.7355\times 10^{-3}$ \\
\hline
$3$ & $0.316259-0.0792091i$ & $1.2974\times 10^{-3}$ & $0.389005-0.0810150i$
& $1.0412\times 10^{-3}$ & $0.398423-0.0837001i$ & $9.9156\times 10^{-4}$ \\
\hline
$4$ & $0.316323-0.0791932i$ & $3.7361\times 10^{-5}$ & $0.389121-0.0809908i$
& $5.9507\times 10^{-5}$ & $0.398446-0.0836952i$ & $1.3216\times 10^{-5}$ \\
\hline
$5$ & $0.316314-0.0791584i$ & $1.8567\times 10^{-5}$ & $0.389122-0.0809964i$
& $3.5878\times 10^{-6}$ & $0.398443-0.0836835i$ & $1.0118\times 10^{-5}$ \\
\hline
$6$ & $0.316323-0.0791561i$ & $5.8056\times 10^{-6}$ & $0.389118-0.0809973i$
& $2.3308\times 10^{-6}$ & $0.398427-0.0836869i$ & $9.5301\times 10^{-6}$ \\
\hline
$7$ & $0.316325-0.0791622i$ & $3.1595\times 10^{-6}$ & $0.389118-0.0809956i$
& $9.0864\times 10^{-7}$ & $0.398425-0.0836773i$ & $4.9112\times 10^{-6}$ \\
\hline
$8$ & $0.316324-0.0791623i$ & $5.5403\times 10^{-6}$ & $0.389118-0.0809955i$
& $3.0076\times 10^{-7}$ & $0.398425-0.0836775i$ & $1.8531\times 10^{-6}$ \\
\hline
$9$ & $0.316327-0.0791731i$ & $1.3535\times 10^{-5}$ & $0.389118-0.0809957i$
& $1.1575\times 10^{-7}$ & $0.398424-0.0836740i$ & $1.7973\times 10^{-6}$ \\
\hline
$10$ & $0.316303-0.0791791i$ & $1.8908\times 10^{-5}$ & $0.389118-0.0809957i$
& $4.7381\times 10^{-8}$ & $0.398424-0.0836740i$ & $1.3883\times 10^{-6}$ \\
\hline
$11$ & $0.316296-0.0791513i$ & $1.4370\times 10^{-5}$ & $0.389118-0.0809956i$
& $1.8440\times 10^{-8}$ & $0.398423-0.0836713i$ & $1.4479\times 10^{-6}$ \\
\hline
$12$ & $0.316298-0.0791507i$ & $3.7402\times 10^{-5}$ & $0.389118-0.0809956i$
& $6.0383\times 10^{-9}$ & $0.398424-0.0836711i$ & $1.5504\times 10^{-6}$ \\
\hline\hline
\end{tabular}%
\vspace{0.2cm}

Table $V$: The fundamental modes of gravitational, electromagnetic, and
scalar perturbations calculated with the WKB formula of different orders.
The calculated frequency is related to the last line of table $I$.

\begin{tabular}{|c|c|c|c|c|c|c|}
\hline\hline
$k$ & $\omega _{k}$ $\left( s=2\right) $ & $\Delta _{k}$ $\left( s=2\right) $
& $\omega _{k}$ $\left( s=1\right) $ & $\Delta _{k}$ $\left( s=1\right) $ & $%
\omega _{k}$ $\left( s=0\right) $ & $\Delta _{k}$ $\left( s=0\right) $ \\
\hline\hline
$2$ & $0.318169-0.244191i$ & $3.9606\times 10^{-2}$ & $0.389204-0.249296i$ &
$3.6355\times 10^{-2}$ & $0.399316-0.256610i$ & $3.7084\times 10^{-2}$ \\
\hline
$3$ & $0.313179-0.237652i$ & $4.1256\times 10^{-3}$ & $0.385440-0.243377i$ &
$3.5161\times 10^{-3}$ & $0.396146-0.251650i$ & $2.9431\times 10^{-3}$ \\
\hline
$4$ & $0.313727-0.237237i$ & $3.4444\times 10^{-4}$ & $0.385882-0.243098i$ &
$2.6218\times 10^{-4}$ & $0.396174-0.251632i$ & $7.3453\times 10^{-5}$ \\
\hline
$5$ & $0.313757-0.237277i$ & $3.8045\times 10^{-5}$ & $0.385899-0.243123i$ &
$1.5183\times 10^{-5}$ & $0.396251-0.251753i$ & $8.8714\times 10^{-5}$ \\
\hline
$6$ & $0.313803-0.237242i$ & $5.0312\times 10^{-5}$ & $0.385897-0.243124i$ &
$1.7412\times 10^{-6}$ & $0.396163-0.251809i$ & $5.2559\times 10^{-5}$ \\
\hline
$7$ & $0.313853-0.237307i$ & $4.6614\times 10^{-5}$ & $0.385895-0.243122i$ &
$1.7042\times 10^{-6}$ & $0.396168-0.251817i$ & $1.6805\times 10^{-5}$ \\
\hline
$8$ & $0.313818-0.237334i$ & $3.3418\times 10^{-5}$ & $0.385897-0.243121i$ &
$1.1457\times 10^{-6}$ & $0.396140-0.251835i$ & $1.6169\times 10^{-5}$ \\
\hline
$9$ & $0.313848-0.237374i$ & $8.4617\times 10^{-5}$ & $0.385898-0.243122i$ &
$5.7540\times 10^{-7}$ & $0.396141-0.251836i$ & $1.0537\times 10^{-5}$ \\
\hline
$10$ & $0.313719-0.237472i$ & $1.2630\times 10^{-4}$ & $0.385897-0.243122i$
& $2.5460\times 10^{-7}$ & $0.396124-0.251847i$ & $1.0515\times 10^{-5}$ \\
\hline
$11$ & $0.313602-0.237317i$ & $9.7304\times 10^{-5}$ & $0.385897-0.243122i$
& $1.0297\times 10^{-7}$ & $0.396123-0.251846i$ & $1.0891\times 10^{-5}$ \\
\hline
$12$ & $0.313592-0.237324i$ & $3.0615\times 10^{-4}$ & $0.385897-0.243122i$
& $3.4979\times 10^{-8}$ & $0.396105-0.251858i$ & $1.1112\times 10^{-5}$ \\
\hline\hline
\end{tabular}%
\vspace{0.2cm}

Table $VI$: The first overtone frequency of gravitational,
electromagnetic, and scalar perturbations calculated with the WKB
formula of different orders. The calculated frequency is related
to the last line of table $III$.
\end{center}

\section{Conclusions \label{Conclusions}}

We have investigated the effects of both the axial gravitational and
electromagnetic perturbations on a black hole system in Weyl gravity. We
have derived\ the master equation, describing the QN radiation, from the
conformal invariance property of the Weyl action, and also, a relation
between the Schwarzschild-(a)dS black holes and Weyl solutions. We have
found that the QNM spectra of the Weyl solutions deviate from those of the
Schwarzschild black hole due to the presence of a linear $r$-term in the
metric function. We have seen that, unlike the non-singular black holes in
conformal gravity, this deviation was present even in the eikonal regime.
Thus, it will be possible to test the Weyl solutions (or at least find a
constraint on the free parameter $c$ in order to recover the present
universe after a phase transition where the conformal symmetry is broken)
with the help of future gravitational wave detectors. Moreover, it was shown
that the perturbations in the conformal black holes' background live longer
compared to the Schwarzschild ones.

In addition, we have calculated the QN frequencies of scalar,
electromagnetic, and gravitational perturbations through both the
sixth order WKB approximation and the improved AIM after $15$
iterations. For the obtained frequencies, the effective potential
was positive and all the frequencies had a negative imaginary
part. Therefore, one can obtain some stable black hole solutions
in conformal gravity under these kinds of perturbations.

We have found that the QN modes of gravitational perturbations live longer
with lower frequency compared to the scalar and electromagnetic
perturbations. In addition, all kinds of perturbations decay faster with
more oscillations by increasing the free parameters $d$\ and/or $b$.
Furthermore, the time evolution of different perturbations for early and
intermediate times is studied by using the time-domain integration of the
master equation. The time-domain profile of modes confirmed the previous
results mentioned above.

\section{acknowledgements}

We wish to thank Shiraz University Research Council.

\appendix

\section{Gravitational perturbations of black holes conformally related to
the Schwarzschild-(a)dS solutions \label{appendix}}

Assume a black hole spacetime is conformally related to the
Schwarzschild-(a)dS solutions so that
\begin{equation}
ds^{2}=S(\rho )d\tilde{s}^{2}=\hat{g}_{\mu \nu }dx^{\mu }dx^{\nu }.
\label{gline}
\end{equation}

Multiplying the Schwarzschild spacetime by a conformal factor can be
described by an anisotropic fluid with the following effective
energy-momentum tensor \cite{CYChen}
\begin{equation}
T_{\mu \nu }=\left( \hat{\rho}+p_{2}\right) u_{\mu }u_{\nu }+\left(
p_{1}-p_{2}\right) x_{\mu }x_{\nu }+p_{2}\hat{g}_{\mu \nu },  \label{emt}
\end{equation}%
where $p_{1}$ and $p_{2}$ are, respectively, the radial pressure and the
tangential pressure, and $\hat{\rho}$ is the energy density measured by a
comoving observer. The explicit expressions of $\hat{\rho}$, $p_{1}$, and $%
p_{2}$ are functions of $\rho $ and they can be derived by calculating the
corresponding field equations constructed from the line element (\ref{gline}%
). In addition, $u^{\mu }$ is the timelike four-velocity and $x^{\mu }$ is
the spacelike unit vector (orthogonal to $u^{\mu }$ and angular directions)
and they satisfy
\begin{equation}
u_{\mu }u^{\mu }=-1;\ \ \ \ \ \ \ \ x_{\mu }x^{\mu }=1,  \label{ux}
\end{equation}%
in which the indices are raised and lowered by the metric $\hat{g}_{\mu \nu
} $, and we also assumed $u^{\mu }=\left[ u^{t},0,0,0\right] $ and $x^{\mu }=%
\left[ 0,x^{\rho },0,0\right] $ in the comoving frame.

In order to obtain the master wave equation and the effective potential of
the line element (\ref{gline}), we follow Chandrasekhar and his notation is
used \cite{Chandrasekhar}. The axial perturbations are characterized by
introducing the non-vanishing parameters $\sigma $, $q_{2}$, and $q_{3}$ in
the unperturbed spacetime in the following form
\begin{equation}
ds^{2}=-e^{2\nu }dt^{2}+e^{2\psi }\left( d\varphi -\sigma
dt-q_{2}dx^{2}-q_{3}dx^{3}\right) ^{2}+e^{2\mu _{2}}\left( dx^{2}\right)
^{2}+e^{2\mu _{3}}\left( dx^{3}\right) ^{2},  \label{AM}
\end{equation}%
in which $\nu $, $\psi $, $\mu _{2}$, $\mu _{3}$, $\sigma $, $q_{2}$, and $%
q_{3}$ are functions of $t$, $x^{2}$ (radial coordinate $\rho $), and $x^{3}$
(polar angle $\theta $) so that $\sigma $, $q_{2}$, and $q_{3}$ are small
quantities. The components of the unperturbed metric ($\sigma ,q_{2},q_{3}=0$%
) are as follows
\begin{eqnarray}
e^{\nu } &=&\sqrt{S(\rho )g(\rho )};\ \ e^{\mu _{2}}=\sqrt{\frac{S(\rho )}{%
g(\rho )}};\ \ e^{\mu _{3}}=\rho \sqrt{S(\rho )};\ \ e^{\psi }=\rho \sin
\left( \theta \right) \sqrt{S(\rho )},  \label{Acomp} \\
g\left( \rho \right) &=&1-\frac{2M}{\rho }-\frac{\Lambda \rho ^{2}}{3}.
\label{AMF}
\end{eqnarray}

The proper field equations, describing this unperturbed metric, are
\begin{equation}
G_{\mu \nu }+\Lambda g_{\mu \nu }=T_{\mu \nu },  \label{AFE}
\end{equation}%
where $G_{\mu \nu }$ is the Einstein tensor. Here, following \cite{CYChen},
we use the tetrad formalism to derive the master equation because the axial
components of the perturbed energy-momentum tensor in the tetrad frame also
vanish in this case, and thus, $\hat{\rho}$ , $p_{1}$, and $p_{2}$ have
nothing to do with the master equation. The tetrad basis corresponding to
the line element (\ref{AM}) is defined as follows
\begin{equation}
e_{(a)}^{\mu }=\left(
\begin{array}{cccc}
e^{-\nu } & \sigma e^{-\nu } & 0 & 0 \\
0 & e^{-\psi } & 0 & 0 \\
0 & q_{2}e^{-\mu _{2}} & e^{-\mu _{2}} & 0 \\
0 & q_{3}e^{-\mu _{3}} & 0 & e^{-\mu _{3}}%
\end{array}%
\right) ,
\end{equation}%
where the Greek letters (labelling the columns) are the tensor indices and
Latin letters enclosed in parentheses (labelling the rows) are the tetrad
indices run from $0$ to $3$. Based on this formalism, any vector or tensor
field can be projected onto the tetrad frame to obtain its tetrad components
\begin{equation}
\left\{
\begin{array}{c}
A_{(a)}=e_{(a)}^{\mu }A_{\mu };\ \ A^{(a)}=\eta ^{(a)(b)}A_{(b)};\ \ A^{\mu
}=e_{(a)}^{\mu }A^{(a)}, \\
T_{(a)(b)}=e_{(a)}^{\mu }e_{(b)}^{\nu }T_{\mu \nu };\ \ T_{\mu \nu }=e_{\mu
}^{(a)}e_{\nu }^{(b)}T_{(a)(b)},%
\end{array}%
\right.
\end{equation}%
in which $\eta _{(a)(b)}=e_{(a)}^{\mu }e_{\mu (b)}$ is a constant symmetric
matrix. In the tetrad frame, the perturbed field equations (\ref{AFE}) read
\begin{equation}
\delta G_{(a)(b)}+\Lambda \eta _{(a)(b)}=\delta T_{(a)(b)},
\end{equation}
where the perturbed energy-momentum tensor is as follows
\begin{eqnarray}
\delta T_{(a)(b)} &=&\left( \hat{\rho}+p_{2}\right) \delta \left(
u_{(a)}u_{(b)}\right) +\left( \delta \hat{\rho}+\delta p_{2}\right)
u_{(a)}u_{(b)}  \notag \\
&&+\left( p_{1}-p_{2}\right) \delta \left( x_{(a)}x_{(b)}\right) +\left(
\delta p_{1}-\delta p_{2}\right) x_{(a)}x_{(b)}+\delta p_{2}\eta _{(a)(b)}.
\end{eqnarray}

By considering the constraints (\ref{ux}) and $u_{\mu }x^{\mu }=0$, we find
that the axial components of the perturbed energy-momentum tensor in the
tetrad frame vanish
\begin{equation}
\delta T_{(1)(0)}=\delta T_{(1)(2)}=\delta T_{(1)(3)}=0.
\end{equation}

Since the axial components of the perturbed energy-momentum tensor vanish,
the master equation of the axial perturbations can be derived from
\begin{equation}
\delta G_{(a)(b)}+\Lambda \eta _{(a)(b)}=0.  \label{fe}
\end{equation}
By substituting the line element (\ref{AM}) into (\ref{fe}), one can find $%
(1,2)$ and $(1,3)$ components as
\begin{equation}
\left[ e^{3\psi +\nu -\mu _{2}-\mu _{3}}Q_{23}\right] _{,3}=e^{3\psi -\nu
-\mu _{2}+\mu _{3}}\left( q_{2,0}-\sigma _{,2}\right) _{,0},
\end{equation}%
\begin{equation}
\left[ e^{3\psi +\nu -\mu _{2}-\mu _{3}}Q_{23}\right] _{,2}=-e^{3\psi -\nu
+\mu _{2}-\mu _{3}}\left( q_{3,0}-\sigma _{,3}\right) _{,0},
\end{equation}%
where $x^{0}=t$ and $x^{1}=\varphi $, and we used $Q_{AB}=q_{A,B}-q_{B,A}$.
Considering the unperturbed values of $\psi $\ and $\mu _{3}$\ from (\ref%
{Acomp}), we obtain the pair of equations
\begin{equation}
\frac{e^{\nu +\mu _{2}}}{S^{2}\left( \rho \right) \rho ^{4}\sin ^{3}\left(
\theta \right) }\frac{\partial Q}{\partial \theta }=\left( q_{2,0}-\sigma
_{,2}\right) _{,0},  \label{e1}
\end{equation}%
\begin{equation}
\frac{e^{\nu -\mu _{2}}}{S\left( \rho \right) \rho ^{2}\sin ^{3}\left(
\theta \right) }\frac{\partial Q}{\partial \rho }=-\left( q_{3,0}-\sigma
_{,3}\right) _{,0},  \label{e2}
\end{equation}%
which we used the new variable $Q$ with the definition%
\begin{equation}
Q=Q\left( t,\rho ,\theta \right) =e^{3\psi +\nu -\mu _{2}-\mu _{3}}Q_{23}.
\end{equation}

By differentiating Eqs. (\ref{e1}) and (\ref{e2}) and eliminating $\sigma $,
we obtain%
\begin{equation}
\frac{e^{\nu +\mu _{2}}}{S^{2}\left( \rho \right) \rho ^{4}}\frac{\partial }{%
\partial \theta }\left( \frac{1}{\sin ^{3}\left( \theta \right) }\frac{%
\partial Q}{\partial \theta }\right) +\frac{1}{\sin ^{3}\left( \theta
\right) }\frac{\partial }{\partial \rho }\left( \frac{e^{\nu -\mu _{2}}}{%
S\left( \rho \right) \rho ^{2}}\frac{\partial Q}{\partial \rho }\right) =%
\frac{1}{S\left( \rho \right) \rho ^{2}\sin ^{3}\left( \theta \right) e^{\nu
-\mu _{2}}}\frac{\partial ^{2}Q}{\partial t^{2}}.  \label{e4}
\end{equation}

We now consider the following decomposition%
\begin{equation}
Q=e^{-i\omega t}Q\left( \rho \right) C_{\ell +2}^{-3/2}\left( \theta \right)
,  \label{e5}
\end{equation}%
where $C_{\ell +2}^{-3/2}\left( \theta \right) $\ is the Gegenbauer function
governed by the following differential equation%
\begin{equation}
\left[ \frac{d}{d\theta }\left( \sin ^{2\alpha }\left( \theta \right) \frac{d%
}{d\theta }\right) +n\left( n+2\alpha \right) \sin ^{2\alpha }\left( \theta
\right) \right] C_{n}^{\alpha }\left( \theta \right) =0.  \label{e6}
\end{equation}

Using (\ref{e5}) and (\ref{e6}), equation (\ref{e4})\ reduces to%
\begin{equation}
\frac{d}{d\rho }\left( \frac{e^{\nu -\mu _{2}}}{S\left( \rho \right) \rho
^{2}}\frac{dQ}{d\rho }\right) +\left( \frac{\omega ^{2}}{S\left( \rho
\right) \rho ^{2}e^{\nu -\mu _{2}}}-\frac{e^{\nu +\mu _{2}}L}{S^{2}\left(
\rho \right) \rho ^{4}}\right) Q=0,
\end{equation}%
in which $L=\left( \ell -1\right) \left( \ell +2\right) $. By introducing $%
Q=Z\Psi ^{(-)}$, one can find that this equation converts to%
\begin{equation}
\frac{d^{2}\Psi ^{(-)}}{d\rho _{\ast }^{2}}+\left( \omega ^{2}-g\left( \rho
\right) \left[ \frac{\ell \left( \ell +1\right) }{\rho ^{2}}-\frac{2}{\rho
^{2}}-Z\frac{d}{d\rho }\left( \frac{\ g\left( \rho \right) }{Z^{2}}\frac{dZ}{%
d\rho }\right) \right] \right) \Psi ^{(-)}=0,  \label{Amwe}
\end{equation}%
where $Z=\rho \sqrt{S(\rho )}$\ and $\rho _{\ast }=\int e^{-\nu +\mu
_{2}}d\rho =\int g^{-1}\left( \rho \right) d\rho $\ is the tortoise
coordinate. Therefore, this is the master wave equation for axial
gravitational perturbations of black holes conformally related to the
Schwarzschild-adS spacetime by the conformal factor $S(\rho )$. One may note
that for $S(\rho )=1$\ (or $Z=\rho $), this equation reduces to%
\begin{equation}
\frac{d^{2}\Psi ^{(-)}}{d\rho _{\ast }^{2}}+\left[ \omega ^{2}-g\left( \rho
\right) \left( \frac{\ell \left( \ell +1\right) }{\rho ^{2}}-\frac{6M}{\rho
^{3}}\right) \right] \Psi ^{(-)}=0,
\end{equation}%
which is the wave equation of axial perturbations of the Schwarzschild-adS
black holes \cite{SchwarzschildadS}, as it should be.

\section{ELECTROMAGNETIC PERTURBATIONS of Weyl gravity \label{app}}

Here, we consider the evolution of the Maxwell field in Weyl gravity with
the line element (\ref{CS}). The evolution is governed by Maxwell equations%
\begin{equation}
\nabla _{\mu }F^{\mu \nu }=0;\ \ \ F_{\mu \nu }=A_{\nu ,\mu }-A_{\mu ,\nu },
\label{Max}
\end{equation}%
where $F_{\mu \nu }$\ is the Faraday tensor and $A_{\mu }$\ is the
electromagnetic potential. The four-potential $A_{\mu }$ can be expanded in
4-dimensional vector spherical harmonics as \cite{SchwarzschildadS}
\begin{equation}
A_{\mu }\left( t,r,\theta ,\varphi \right) =\sum_{\ell ,m}\left( \left[
\begin{array}{c}
0 \\
0 \\
\frac{a(t,r)}{\sin \left( \theta \right) }\partial _{\varphi }Y_{\ell
m}\left( \theta ,\varphi \right) \\
-a\left( t,r\right) \sin \left( \theta \right) \partial _{\theta }Y_{\ell
m}\left( \theta ,\varphi \right)%
\end{array}%
\right] +\left[
\begin{array}{c}
f(t,r)Y_{\ell m}\left( \theta ,\varphi \right) \\
h(t,r)Y_{\ell m}\left( \theta ,\varphi \right) \\
k(t,r)\partial _{\theta }Y_{\ell m}\left( \theta ,\varphi \right) \\
k(t,r)\partial _{\varphi }Y_{\ell m}\left( \theta ,\varphi \right)%
\end{array}%
\right] \right) ,
\end{equation}%
where $Y_{\ell m}\left( \theta ,\varphi \right) $\ denotes the spherical
harmonics. The first term in the right-hand side has parity $\left(
-1\right) ^{\ell +1}$ (axial sector of the expansion) and the second term
has parity $\left( -1\right) ^{\ell }$ (polar sector of the expansion). By
substituting this expansion into the Maxwell equations (\ref{Max}), one can
find a second-order differential equation for the radial part as (see \cite%
{Cosimo}\ for details of calculations)%
\begin{equation}
\frac{d^{2}\Psi \left( r_{\ast }\right) }{dr_{\ast }^{2}}+\left[ \omega
^{2}-V_{e}\left( r_{\ast }\right) \right] \Psi \left( r_{\ast }\right) =0,
\end{equation}%
\begin{equation}
V_{e}\left( r_{\ast }\right) =f\left( r\right) \frac{\ell \left( \ell
+1\right) }{r^{2}},
\end{equation}%
for both axial and polar sectors, and $r_{\ast }=\int f^{-1}\left( r\right)
dr$\ being the tortoise coordinate. The mode $\Psi \left( r_{\ast }\right) $
is a linear combination of the functions $a(t,r)$, $f(t,r)$, $h(t,r)$, and $%
k(t,r)$, but a different functional dependence based on the parity; for
axial sector the mode is given by $\Psi \left( r_{\ast }\right) =a(t,r)$
whereas for polar sector it is $\Psi \left( r_{\ast }\right) =\frac{r^{2}}{%
\ell (\ell +1)}\left[ \partial _{t}h(t,r)-\partial _{r}f(t,r)\right] $.


\begin{thebibliography}{99}
\bibitem{AbbottBH} B. P. Abbott et al. (LIGO Scientific and Virgo
Collaborations), Phys. Rev. Lett. 116 (2016) 061102.

\bibitem{Akiyama1} K. Akiyama et al. (Event Horizon Telescope), Astrophys.
J. 875 (2019) L1.

\bibitem{Akiyama4} K. Akiyama et al. (Event Horizon Telescope), Astrophys.
J. 875 (2019) L4.

\bibitem{Bach} R. Bach, Math. Z. 9 (1921) 110.

\bibitem{Buchdahl} A. Buchdahl, Edinburgh Math. Soc. Proc. 10 (1953) 16.

\bibitem{Riegert} R. J. Riegert, Phys. Rev. Lett. 53 (1984) 315.

\bibitem{Mannheim} P. D. Mannheim and D. Kazanas, Astrophys. J. 342 (1989)
635.

\bibitem{Pope} H. Lu and C. N. Pope, Phys. Rev. Lett. 106 (2011) 181302.

\bibitem{MannheimDavidson} P. D. Mannheim and A. Davidson, [arXiv:0001115].

\bibitem{Bender} C. M. Bender, Rep. Prog. Phys. 70 (2007) 947.

\bibitem{BenderMannheim} C. M. Bender and P. D. Mannheim, Phys. Rev. D 78
(2008) 025022.

\bibitem{BenderMannheim2008} C. M. Bender and P. D. Mannheim, J. Phys. A 41
(2008) 304018.

\bibitem{BenderMannheimPRL} C. M. Bender and P. D. Mannheim, Phys. Rev.
Lett. 100 (2008) 110402.

\bibitem{Mannheim2013} P. D. Mannheim, Philos. Trans. Royal Soc. A 371
(2013) 20120060.

\bibitem{MannheimPLB} P. D. Mannheim, Phys. Lett. B 753 (2016) 288.

\bibitem{Mannheim2018} P. D. Mannheim, J. Phys. A 51 (2018) 315302.

\bibitem{Bergshoeff} E. Bergshoeff, M. de Roo and B. de Wit, Nucl. Phys. B
182 (1981) 173.

\bibitem{Wit} B. de Wit, J. W. van Holten and A. Van Proeyen, Nucl. Phys. B
184 (1981) 77.

\bibitem{Stelle} K. Stelle, Phys. Rev. D 16 (1977) 953.

\bibitem{Faria} F. Faria, Eur. Phys. J. C 76 (2016) 188.

\bibitem{Witten} N. Berkovits and E. Witten, JHEP 08 (2004) 009.

\bibitem{Tseytlin} H. Liu and A. A. Tseytlin, Nucl. Phys. B 533 (1998) 88.

\bibitem{Henningson} M. Henningson and K. Skenderis, JHEP 07 (1998) 023.

\bibitem{Balasubramanian} V. Balasubramanian, E. G. Gimon, D. Minic and J.
Rahmfeld, Phys. Rev. D 63 (2001) 104009.

\bibitem{Adler} S. L. Adler, Rev. Mod. Phys. 54 (1982) 729.

\bibitem{Hooft} G. 't Hooft, Found. Phys. 41 (2011) 1829.

\bibitem{Maldacena} J. Maldacena, [arXiv:1105.5632].

\bibitem{Anastasiou} G. Anastasiou and R. Olea, Phys. Rev. D 94 (2016)
086008.

\bibitem{AbbottNS} B. Abbott et al. (LIGO Scientific and Virgo
Collaborations), Phys. Rev. Lett. 119 (2017) 161101.

\bibitem{Blanchet} L. Blanchet, [arXiv:1902.09801].

\bibitem{Pretorius} F. Pretorius, Phys. Rev. Lett. 95 (2005) 121101.

\bibitem{Campanelli} M. Campanelli, C. O. Lousto, P. Marronetti and Y.
Zlochower, Phys. Rev. Lett. 96 (2006) 111101.

\bibitem{Baker} J. G. Baker, J. Centrella, D. I. Choi, M. Koppitz and J.
vanMeter, Phys. Rev. Lett. 96 (2006) 111102.

\bibitem{BertiCardosoWill} E. Berti, V. Cardoso and C. M. Will, Phys. Rev. D
73 (2006) 064030.

\bibitem{Barack} L. Barack et al., Class. Quant. Grav. 36 (2019) 143001.

\bibitem{Regge} T. Regge and J. A. Wheeler, Phys. Rev. 108 (1957) 1063.

\bibitem{Zerilli} F. J. Zerilli, Phys. Rev. D 2 (1970) 2141.

\bibitem{Mashhoon} V. Ferrari and B. Mashhoon, Phys. Rev. D 30 (1984) 295.

\bibitem{Schutz} B. F. Schutz and C. M. Will, Astrophys. J. Lett. 291 (1985)
L33.

\bibitem{Iyer} S. Iyer and C. M. Will, Phys. Rev. D 35 (1987) 3621.

\bibitem{KonoplyaWKB} R. A. Konoplya, Phys. Rev. D 68 (2003) 024018.

\bibitem{ChandrasekharDetweiler} S. Chandrasekhar and S. Detweiler, Proc. R.
Soc. London, Ser. A 344 (1975) 441.

\bibitem{Leaver} E. W. Leaver, Proc. R. Soc. London, Ser. A, 402 (1985) 285.

\bibitem{Horowitz} G. T. Horowitz and V. E. Hubeny, Phys. Rev D 62 (2000)
024027.

\bibitem{Naylor} H. T. Cho et al., Adv. Math. Phys. 2012 (2012) 281705.

\bibitem{Hatsuda} Y. Hatsuda, [arXiv:1906.07232].

\bibitem{Kokkotas} K. D. Kokkotas and B. G. Schmidt, Living Rev. Rel. 2
(1999) 2.

\bibitem{Berti} E. Berti, V. Cardoso and A. O. Starinets, Class. Quant.
Grav. 26 (2009) 163001.

\bibitem{Konoplya} R. A. Konoplya and A. Zhidenko, Rev. Mod. Phys. 83 (2011)
793.

\bibitem{EG} V. Cardoso, R. Konoplya and J. P. S. Lemos, Phys. Rev. D 68
(2003) 044024.

\bibitem{Manfredi} L. Manfredi, J. Mureika, J. Moffat, Phys. Lett. B 779
(2018) 492.

\bibitem{RongjiaYang} R. Yang, Phys. Lett. B 784 (2018) 212.

\bibitem{Caprini} C. Caprini, P. Holscher and D. J. Schwarz, Phys. Rev. D 98
(2018) 084002.

\bibitem{Holscher} P. Holscher and D. J. Schwarz, Phys. Rev. D 99 (2019)
084005.

\bibitem{FariaPRD} F. F. Faria, Phys. Rev. D 99 (2019) 048501.

\bibitem{Cosimo} B. Toshmatov, C. Bambi, B. Ahmedov, Z. Stuchlik and J.
Schee, Phys. Rev. D 96 (2017) 064028.

\bibitem{CYChen} C. Y. Chen and P. Chen, Phys. Rev. D 99 (2019) 104003.

\bibitem{Hendi} S. H. Hendi, M. Momennia and F. Soltani Bidgoli,
[arXiv:1807.01792].

\bibitem{Momennia} M. Momennia and S. H. Hendi, Phys. Rev. D 99 (2019)
124025.

\bibitem{PangPope} H. Lu, Y. Pang, C. N. Pope and J. F. Vazquez-Poritz,
Phys. Rev. D 86 (2012) 044011.

\bibitem{SchwarzschildadS} V. Cardoso and J. P. S. Lemos, Phys. Rev. D 64
(2001) 084017.

\bibitem{ZhidenkoDS} A. Zhidenko, Class. Quant. Grav. 21 (2004) 273.

\bibitem{Matyjasek} J. Matyjasek and M. Opala, Phys. Rev. D 96 (2017) 024011.

\bibitem{Ciftci} H. Ciftci, R. L. Hall and N. Saad, J. Phys. A 36 (2003)
11807.

\bibitem{CiftciHall} H. Ciftci, R. L. Hall and N. Saad, Phys. Lett. A 340
(2005) 388.

\bibitem{ChoNaylor} H. T. Cho, A. S. Cornell, J. Doukas and W. Naylor,
Class. Quant. Grav. 27 (2010) 155004.

\bibitem{Pullin} C. Gundlach, R. H. Price and J. Pullin, Phys. Rev. D 49
(1994) 883.

\bibitem{Zhidenko} R. A. Konoplya, A. Zhidenko and A. F. Zinhailo, Class.
Quant. Grav. 36 (2019) 155002.

\bibitem{Chandrasekhar} S. Chandrasekhar, \textit{The Mathematical Theory of
Black Holes} (Oxford University, Oxford, 1983).
\end{thebibliography}
\end{document}